\documentclass{aa}

\usepackage{txfonts}
\usepackage{subcaption}         
\usepackage{lscape}             
\usepackage{placeins}           

\usepackage{CJK}
\usepackage{graphicx}
\usepackage{amsmath,amssymb}

\usepackage{dblfloatfix}

\usepackage{xcolor}

\begin{document}

\title{Asymmetry in the protostellar system HOPS 198}
\subtitle{Evidence for the evolution of outflow opening angle driven by density of the surrounding core}
\authorrunning{Donglin Wu et al.}

\author{Donglin Wu \begin{CJK*}{UTF8}{gbsn}(吴东霖)\end{CJK*}\inst{1}\fnmsep\thanks{Corresponding author: donglin.wu@yale.edu} 
\and H\'ector Arce \inst{1}
\and Cheng-Han Hsieh \begin{CJK*}{UTF8}{bsmi}(謝承翰)\end{CJK*} \inst{1}\inst{2}\fnmsep \thanks{The NASA Hubble Fellowship Program Sagan Fellow.}} 
\institute{Department of Astronomy, Yale University, New Haven, CT 06511, USA
\and Department of Astronomy, The University of Texas at Austin, 2515 Speedway, Stop C1400, Austin, TX 78712-1205, USA
} 

\date{}

\abstract
{Protostellar outflows are thought to be responsible for the low star formation efficiency of protostellar cores. However, whether outflows can disperse a significant fraction of the gas in the core depends on the outflow opening angle. It is established that the outflow opening angle increases during the early stages of the protostellar evolution, but the underlying mechanism is poorly understood.
Observations of HOPS 198, a Class 0 protostar in the Orion A molecular cloud, provide insights into this question.
HOPS 198 exhibits a strong east-west asymmetry in its outflow and its core. The opening angle of the eastern lobe ($\sim80^{\circ}$) is more than twice wider than that of the western lobe ($\sim30^{\circ}$), while the surface density of the west side of the core is $1.5-2.8$ times higher than the east side. 
Using an analytical model in which the molecular outflow morphology is shaped by interactions between the wide-angle protostellar wind ($\gtrsim 80^{\circ}$) and surrounding material in the core, we find that the difference in opening angle for the two lobes can be explained by the difference in core density on the two sides. This result supports the hypothesis that the evolution of the outflow opening angle is driven by the evolution in the density of the protostellar core.}


\maketitle

\nolinenumbers

\section{Introduction} 
\label{sec:intro}
Protostellar outflows play a crucial role in star formation. On parsec scales, they are thought to drive turbulence in regions of active cluster formation \citep{Matzner2007, Nakamura2007}, which regulates subsequent star formation. At smaller scales, individual  outflows disperse the gas of  their surrounding  protostellar core and in this way modulate the infall of material onto the protostar. This might also affect the final mass of the star. 
In particular, outflows have been proposed to explain the low star formation efficiency in cores, where studies have suggested that only $\sim 30\%$ of the mass in the core ends up in the stars \citep{Matzner2000, Evans2009, Hansen2012, Zhang2016}.

However, it remains an open question whether outflows are capable of dispersing approximately $70\%$ of the mass. One important factor to be considered is the outflow opening angle. Based on simulations, it was proposed that wide-angle outflows have higher mass-loss rates and are more effective in removing materials from the core than collimated jets \citep{Matzner2000, Machida2008, Machida2012, Offner2014}. Observations also suggested that wide-angle outflows have sufficient momentum and energy to disperse material from the core (e.g., \citealt{Arce2010, Dunham2014, Hsieh2023}). 

Therefore, to understand whether outflows cause the low star formation efficiency in cores, the evolution of the outflow opening angles over the lifetime of protostars has to be characterized. 
\citet{Arce2006} found that the opening angle of outflow cavities increases as protostars evolve from Class 0 sources to Class I and Class II sources. This was supported by hydrodynamical simulations \citep{Offner2011}. More recent observational studies have confirmed an increase in the outflow opening angle as sources evolve from  Class 0 to Class I protostars and also found a plateau in the opening angle when the protostars reached the Class I phase. These include studies based on \textit{Spitzer} and other infrared observations \citep{Velusamy2014, Hsieh2017} as well as studies that trace outflows with molecular transitions of CO isotopologs \citep{Hsieh2023, Dunham2024}.

Although the increase in the opening angle over time is relatively well established, it remains uncertain what drives this evolution. Outflows have been understood as gas entrained by protostellar winds that are launched from the protostar or its disk. It was proposed that the protostellar wind consists of both a collimated jet and a wide-angle wind \citep{Kwan1995, Konigl2000, Shu2000}. This was supported by  observations (e.g., \citealt{Bacciotti2000, Arce2002, Arce2013, Zapata2014, Tabone2017, Lee2021}). \citet{Arce2006} argued that the evolution of the opening angle can be accounted for by this two-component stellar wind scenario: only the dense collimated jet can break out of the dense infalling envelope during the early stages of protostellar evolution, and the wide-angle wind can break through only when the envelope has lost enough mass through infall and dispersal. 

We have found potential evidence that the evolution in the outflow opening angle is driven by the outflow-core interaction, particularly the evolution in the protostellar core density. The evidence comes from recent molecular line observations of the environment around the young protostar HOPS 198 using the Atacama Large Millimeter/submillimeter Array telescope (ALMA). 
HOPS 198 is a Class 0 protostar with a bolometric temperature and  luminosity of  $61.4$ K and $0.85~L_{\odot}$ \citep{Furlan2016}. It is located in the southern part of the Orion A cloud in the North LDN 1641 region and lies at a distance of $386$ pc \citep{Tobin2020}.

We show that the molecular core around HOPS 198 and its molecular outflow are strongly asymmetric: the two lobes of the molecular outflow differ significantly in their opening angles, and the core shows a strong density gradient across the protostellar disk axis. 
Using an analytical model that describes the shape of the molecular outflow assuming pressure balance between the core and the protostellar wind, we show that the difference in outflow opening angle can be explained by the difference in core density. This result directly links the ambient core density to the outflow morphology and supports the interpretation in which a declining core density leads to progressively wider outflow opening angles.

This paper is organized as follows. In Section \ref{sec:obs} we describe the ALMA observations. We examine the asymmetries in the surface density map of the core and the outflow in Section \ref{sec:asym_core} and \ref{sec:asym_out}, and we discuss the implications of the results on the evolution of the outflow opening angle in Section \ref{sec:oa_evol}.

\section{Observations}
\label{sec:obs}
HOPS 198 was observed by ALMA during cycle 6 (project ID: 2018.1.00744.S, PI: H. Arce). The observations were conducted in Band 6 and included the 1.3 mm dust continuum emission along with the following six molecular lines: $^{12}\mathrm{CO}(2\!-\!1),\ ^{13}\mathrm{CO}(2\!-\!1),\ \mathrm{C}^{18}\mathrm{O}(2\!-\!1),\ \mathrm{H}_2\mathrm{CO}(3_{0,3}\!-\!2_{0,2}),\ \mathrm{SiO}(5\!-\!4)$, and $\mathrm{N}_2\mathrm{D}^+(3\!-\!2)$.  The details of the observations are described in \citet{Hsieh2023}. The spectral line maps have synthesized beams with dimensions of approximately $1.3^{\prime\prime}\times 1.0^{\prime\prime}$ (or $502~\mathrm{au} \times 386~\mathrm{au}$).
The field of view spans $\sim50^{\prime\prime}$ (or $19300~\mathrm{au}$).
No $\mathrm{SiO}(5\!-\!4)$ or $\mathrm{N}_2\mathrm{D}^+(3\!-\!2)$ emission was detected for HOPS 198. 
For $^{12}\mathrm{CO}(2\!-\!1)$, $^{13}\mathrm{CO}(2\!-\!1)$, $\mathrm{C}^{18}\mathrm{O}(2\!-\!1)$, and $\mathrm{H}_2\mathrm{CO}(3_{0,3}\!-\!2_{0,2})$, the rms noise in the image center is 6.3, 8.3, 6.1, and 5.6 $\mathrm{mJy\,beam^{-1}}$, respectively, and increases by approximately 50$\%$ at about  $\sim20^{\prime\prime}$ from the center of the field of view. The spectral resolution is 0.079, 0.083, 0.083, and 0.084 $\mathrm{km/s}$ for the four lines, respectively.

In addition to the ALMA observations, we used the $\mathrm{C}^{18}\mathrm{O}(1\!-\!0)$ map from the Combined Array for Research in Millimeter Astronomy (CARMA)-Nobeyama Radio Observatory (NRO) Orion survey, which was described in detail in \citet{Kong2018}.
This spectral cube has a synthesized beam of  $10^{\prime\prime}\times 8^{\prime\prime}$ (or $3860~\mathrm{au} \times 3090~\mathrm{au}$), a median rms noise of $0.47$ K,  a spectral resolution of 0.22 $\mathrm{km/s}$, and a  pixel size of $2^{\prime\prime}$ (roughly $772~\mathrm{au}$). 
Unless otherwise specified, all molecular lines mentioned in this paper refer to those obtained by the ALMA observations.

\section{Results and discussions}
\label{sec:res}

\subsection{Overview of the HOPS 198 molecular outflow}
\begin{figure*}
    \centering
    \includegraphics[width=0.99\linewidth]{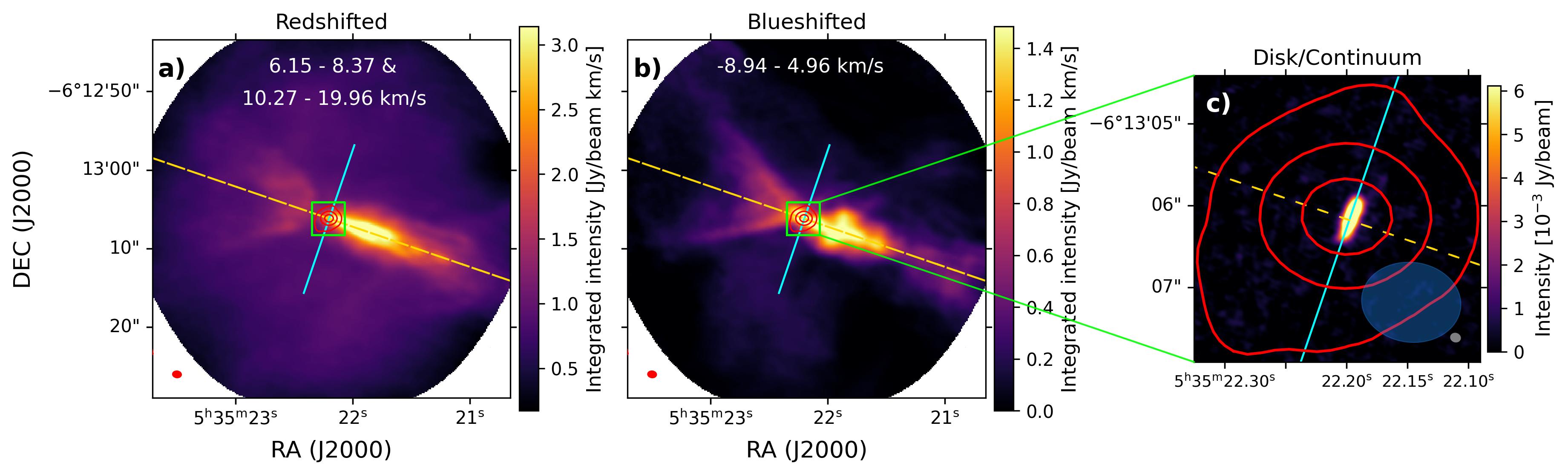}
    \caption{a) Integrated intensity map of $^{12}$CO for redshifted channels with $6.15 \leq v_{\rm los} \leq 8.37$ km/s and $10.27 \leq v_{\rm los} \leq 8.37$ km/s. b) Integrated intensity map of $^{12}$CO for blueshifted channels with $-8.94 \leq v_{\rm los} \leq 4.96$ km/s. c) Zoomed-in continuum image of the disk observed by \citet{Tobin2020} using their ALMA 0.87 mm continuum image. The red contours show the continuum emission from our dataset at levels of $7\sigma$, $40\sigma$, and $142\sigma$, where the rms noise is $\sigma \simeq 0.07$ mJy/beam. 
    The green square in panels (a) and (b) indicates the region that is shown in panel (c). In all panels, the dashed gold line indicates the outflow axis, while the solid cyan line indicates the disk axis. The red ellipses in the bottom left corner of panels (a) and (b) illustrate the beam size of our $^{12}$CO observation. In panel (c), the small gray ellipse and large blue ellipse in the bottom right corner illustrate the beam size for disk observations by \citet{Tobin2020} and for our continuum observation, respectively.
    }
    \label{fig:general}
\end{figure*}

The molecular outflow of HOPS 198 was first observed by \citet{Hsieh2023} as part of a survey. Panels (a) and (b) of Figure \ref{fig:general} show the $^{12}$CO integrated intensity map for the molecular outflow lobes that are red- and blueshifted relative to the system velocity along the line of sight ($v_{\text{sys}} = 5.53$ km/s, \citealt{Hsieh2023}). Channels around the system velocity ($5.03 \leq v_{\text{los}} \leq 6.07$ km/s) were excluded from the integrated intensity map as they are dominated by the emission from the core surrounding the protostar, and channels with $8.45 \leq v_{\text{los}} \leq 10.19$ km/s were excluded in order to avoid including emission from another molecular cloud velocity component unrelated to HOPS 198. 

The two lobes of the outflows are visible in both the blue- and redshifted channels because the source is seen nearly edge-on, with the outflow axis at approximately $80^{\circ}$ with respect to the line of sight \citep{Hsieh2023}, and the two outflow lobes have components of gas moving toward and away from us. The inclination angle with respect to the line of sight ($i$) was derived from the ratio of the deconvolved disk major and minor axes measured by ALMA 0.87 mm continuum data from \citet{Tobin2020}, which is shown in panel (c) of Figure \ref{fig:general}. The maximum line-of-sight velocity of the outflow observed at $3\sigma$ relative to the system velocity is $\sim$10 km/s and $\sim$13 km/s for the blue- and redshifted lobes, respectively. With $i\sim80^{\circ}$, this corresponds to an outflow velocity of at least $\sim$58 km/s and $\sim$75 km/s for the blue- and redshifted lobes, respectively.

Differences in morphology and intensity between the two outflow lobes are visible in the integrated intensity maps. In the red- and blueshifted channels, the eastern lobe exhibits a wide-angle V-shaped morphology with relatively faint emission. In contrast, the western lobe shows brighter and more collimated emissions along the outflow axis. This asymmetry in the outflow morphology is discussed in Section \ref{sec:asym_out} in greater detail, and channel maps for the $^{12}$CO emissions can be found in Appendix \ref{sec:channel_map}.

We determined the disk axis using the 0.87 mm continuum image of the disk. The outflow axis is defined to be perpendicular to the disk axis. The two axes are shown in panel (c) of Figure \ref{fig:general}. In the same panel, we also plot the contours of the dust emission from the continuum image obtained simultaneously with our ALMA line data. The beam of our dataset does not resolve the disk, so the brightest continuum contours appear as ellipses. The $\sim 7 \sigma$ contour is more irregular and appears broader to the south than to the north, possibly tracing emission from the inner envelope.

To analyze the morphology of the core and the outflow further, we computed their masses and surface densities.
We computed the surface density map of the core using the C$^{18}$O line maps from the ALMA and CARMA-NRO observations (Figure \ref{fig:mass_core}) because C$^{18}$O is a good tracer for relatively dense gas in protostellar cores (see Appendix \ref{sec:Herschel} for more information). We computed the molecular outflow surface density using the  $^{12}$CO(2-1) map after cloud subtraction and opacity correction from the ALMA observations (Figure \ref{fig:mass_outflow}). The details of the procedures, including the newly developed technique for the cloud subtraction, are presented in Appendix \ref{sec:mass}.

\begin{figure*}
    \centering
    \includegraphics[width=0.49\linewidth]{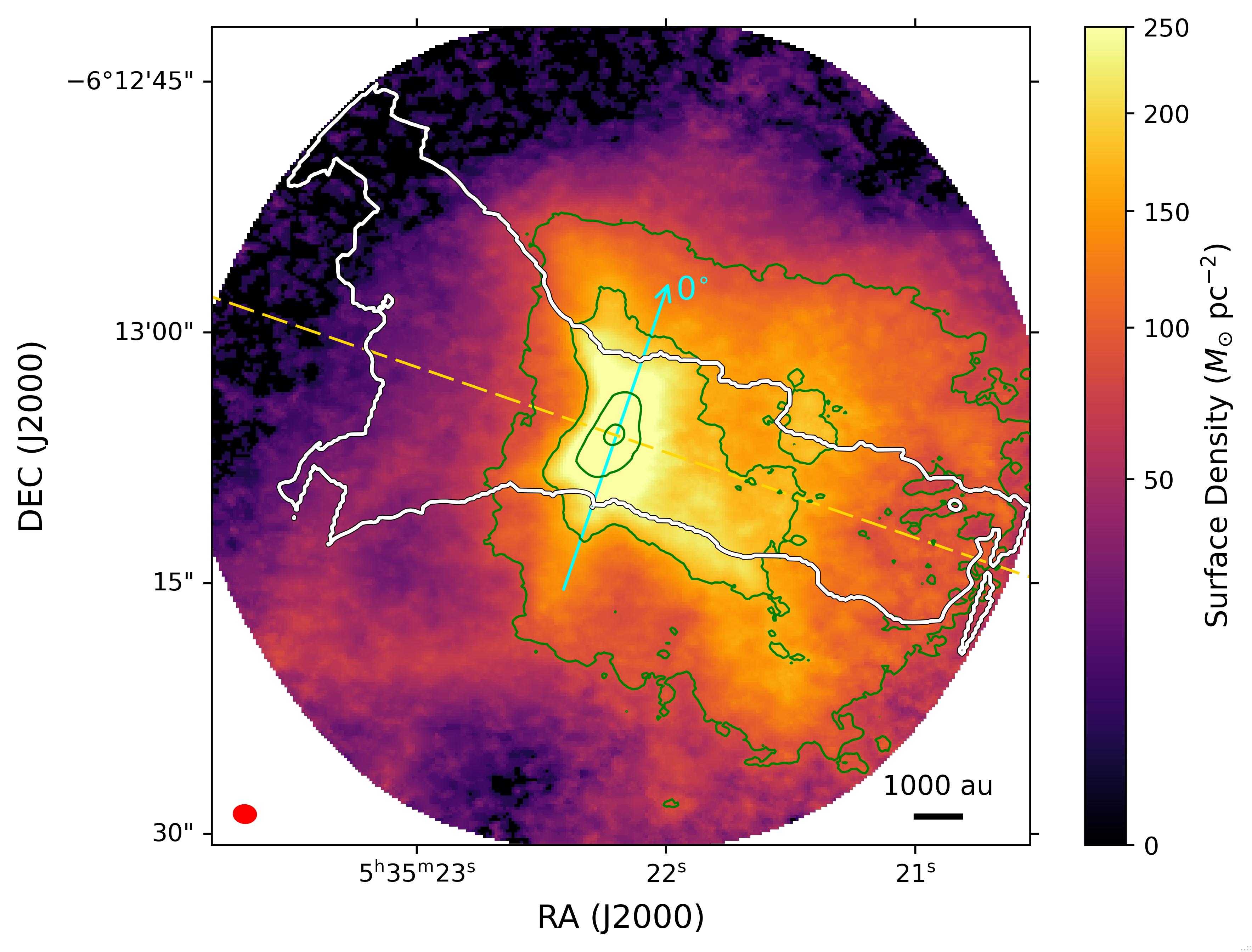}
    \includegraphics[width=0.49\linewidth]{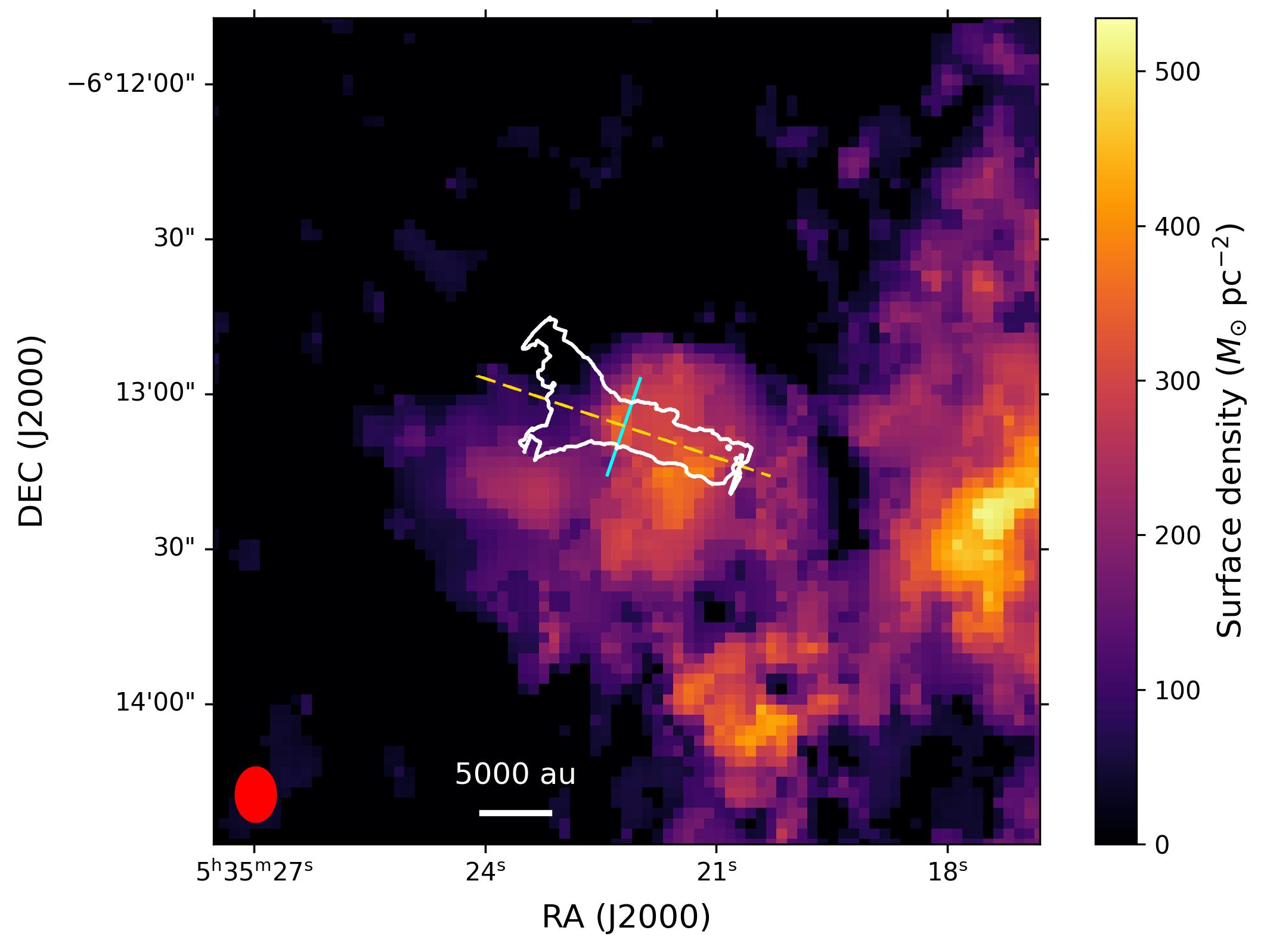}
    \caption{Left: Surface density map of the core computed from the C$^{18}$O (2-1) emission obtained from  ALMA. The color scale uses a square-root stretch and saturates at $250~ M_{\odot}~\mathrm{pc}^{-2}$.
    The green contours indicate $10\%, 20\%, 40\%$, and $80\%$ of the maximum surface density ($850~M_{\odot}~\mathrm{pc}^{-2}$). Right: Surface density map of the core using the CARMA-NRO Orion survey C$^{18}$O (1-0)  data. For both panels, the white contour corresponds to  $3\%$ of the maximum value of the molecular outflow surface density ($106~M_\odot~\mathrm{pc}^{-2}$). The dashed gold line indicates the outflow axis, and the solid cyan arrow is the direction perpendicular to the outflow axis. In the left panel, the arrow points in the direction of 0$^{\circ}$ for the polar distribution of the mass shown in Figure \ref{fig:mass_polar}. The red ellipse in the bottom left corner of each panel indicates the beam size. The linear scales for each panel are shown assuming a distance of 386 pc for HOPS 198. 
    }
    \label{fig:mass_core}
\end{figure*}

\begin{figure}
    \centering
    \includegraphics[width=0.98\linewidth]{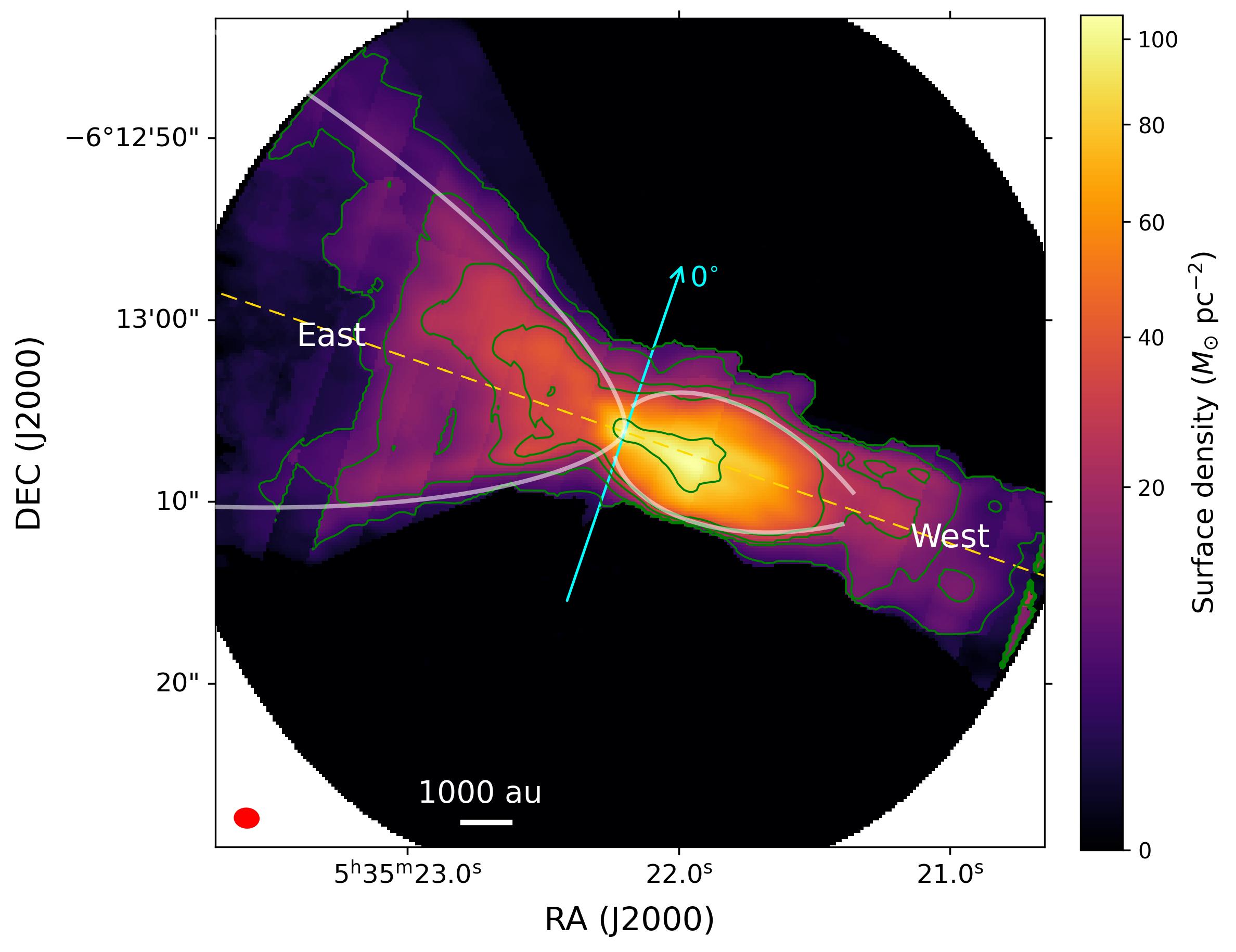}
    \caption{Surface density map of the molecular outflow computed from the opacity-corrected cloud-subtracted $^{12}$CO (2-1) emission observed with ALMA, with green contours indicating $3\%, 10\%, 20\%, 30\%$, and $80\%$ of the maximum value ($106~M_\odot~\mathrm{pc}^{-2}$). The color scale uses a square-root stretch. The white lines represent the shape of the outflow given by Equation \ref{eq:shape_outflow2} using median posterior parameters obtained in Section \ref{sec:shape_out}. The gold and cyan lines are the same as those in Figure \ref{fig:mass_core}.}
    \label{fig:mass_outflow}
\end{figure}

\subsection{The east-west asymmetry in the core}
\label{sec:asym_core}

The east-west asymmetry in the surface density of the core is significant. The left panel of Figure \ref{fig:mass_core} shows the surface density in the core computed from ALMA data. The two outermost green contours, corresponding to $86~M_{\odot}~\mathrm{pc}^{-2}$ and $170~M_{\odot}~\mathrm{pc}^{-2}$, are predominantly in the western lobe. The total mass of the core is estimated to be $0.50 \textrm{ M}_{\odot}$, with $0.16 M_{\odot}$ on the east side of the disk axis, and $0.34 M_{\odot}$ on the west side.
This asymmetry is also obvious on a larger scale with the CARMA-NRO observations, shown in the right panel of Figure \ref{fig:mass_core}. Interestingly, it shows that HOPS 198 is at the edge of the cloud. Most of the gas in the cloud is located in the southwest direction of the protostar.  

\begin{figure}
    \centering
    \includegraphics[width=0.98\linewidth]{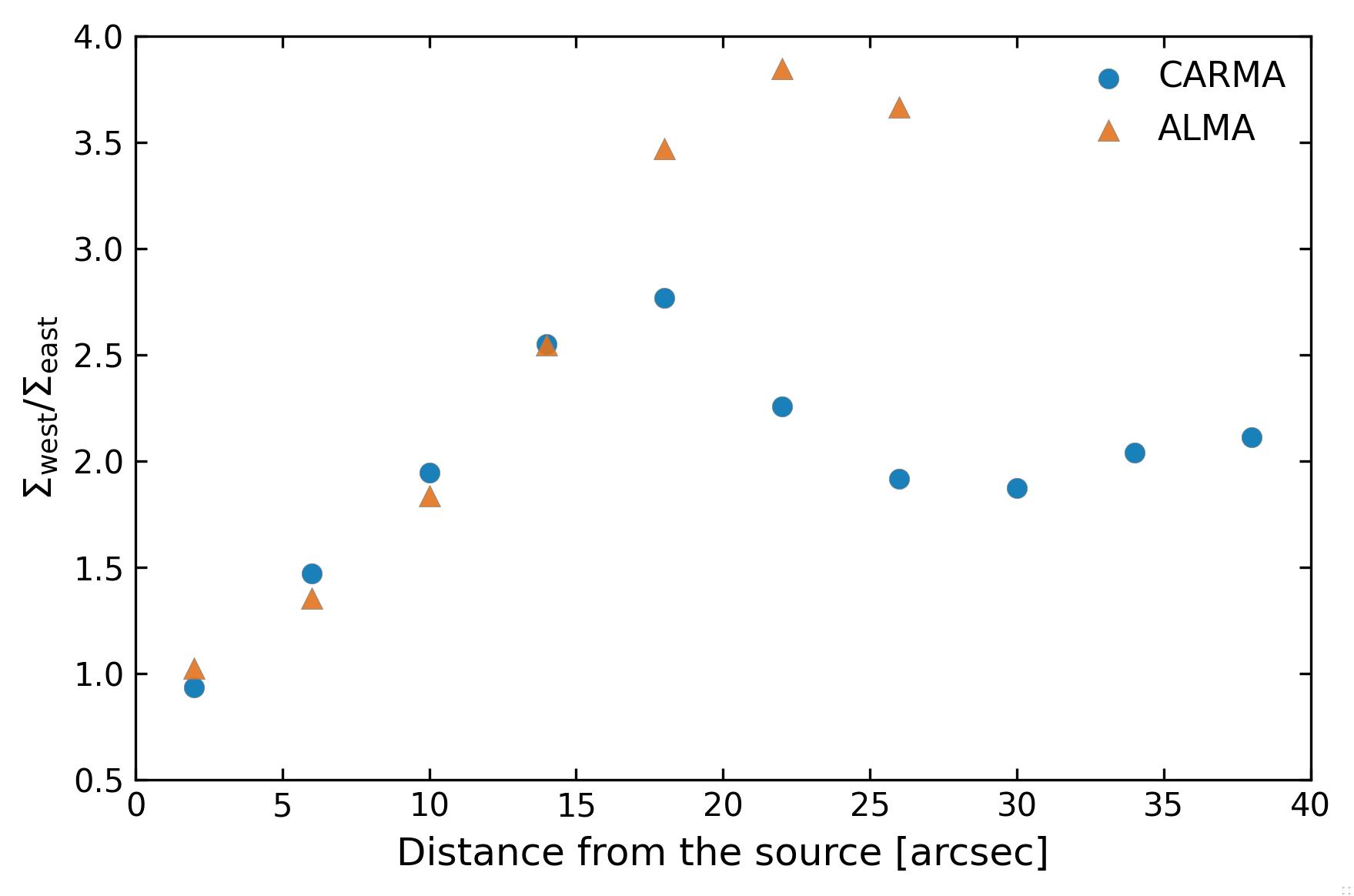}
    \caption{Ratio of the surface density of the west side of the core to the east side in different semicircular annuli with a radial width of 4$^{\prime\prime}$. The triangles and circle represent  estimates obtained from the ALMA and  CARMA-NRO observations, respectively. The distance from the source for each point indicates the mean radius of the annuli from where the surface density ratio was obtained.}
    \label{fig:density_asym}
\end{figure}

To further quantify the asymmetry in surface density, we computed the ratio in the average surface density of the west side of the core to the east side in semicircular annuli with a radial width of 4$^{\prime\prime}$ ($\sim$ 1540 au) using ALMA and CARMA-NRO observations. 
The result is shown in Figure \ref{fig:density_asym}.

The figure shows that the ratios in surface density computed from the ALMA and CARMA-NRO observations agree well for $r \leq 16^{\prime\prime}$. The two sides of the core have approximately the same surface density for $r \leq 4^{\prime\prime}$, which is predominantly the inner envelope. Outside the compact inner envelope, the difference in surface density of the two sides becomes more apparent. The surface density of the western lobe is higher by $50\%$ than the eastern lobe for $4^{\prime\prime} < r \leq 8^{\prime\prime}$, and higher by over $150\%$ than the eastern lobe for $12^{\prime\prime} < r \leq 16^{\prime\prime}$. 

The ratio in surface density computed from ALMA observations continues to rise for $r > 16^{\prime\prime}$. However, this is likely caused by the limited sensitivity at the edge of the ALMA map. The $\mathrm{C}^{18}\mathrm{O}(2\!-\!1)$ emission in the eastern part of the core is too faint 
to be included in our mass estimate (we only used emission above  $3\sigma$; see Appendix \ref{sec:mass}), especially at the periphery of the map, where the sensitivity is lower by 50\% than in the map center.  
The more reliable estimates from CARMA-NRO suggest that the surface density of the western side of the core remains twice higher than that of the east side even at $40^{\prime\prime}$ away from the protostar. This suggests that the asymmetry in the core surface density across the two sides of the core is global.

\subsection{The east-west asymmetry in the outflow}
\label{sec:asym_out}
\subsubsection{Interactions between the outflow and the core}

Figure \ref{fig:mass_outflow} shows the molecular outflow surface density map, which is also strongly asymmetric. The total mass of the outflow is estimated to be $0.037 \textrm{ M}_{\odot}$, with $0.015 \textrm{ M}_{\odot}$ on the east side of the disk axis and $0.022 \textrm{ M}_{\odot}$ on the west side. 
There is a large difference in the surface density between the two lobes because the eastern lobe is wider and covers a larger area in the plane of the sky.  At a distance of 5$^{\prime\prime}$ away from the protostar, the surface density is $25-35~M_{\odot}~\mathrm{pc}^{-2}$ and $70-100~M_{\odot}~\mathrm{pc}^{-2}$ for the eastern and western lobes, respectively.
The difference in the surface density is likely caused by the mass asymmetry in the core: with initially more gas on the west side of the core, the protostellar wind is able to entrain more core gas, making it part of the molecular outflow. 

\begin{figure*}
    \centering
    \includegraphics[width=0.9\linewidth]{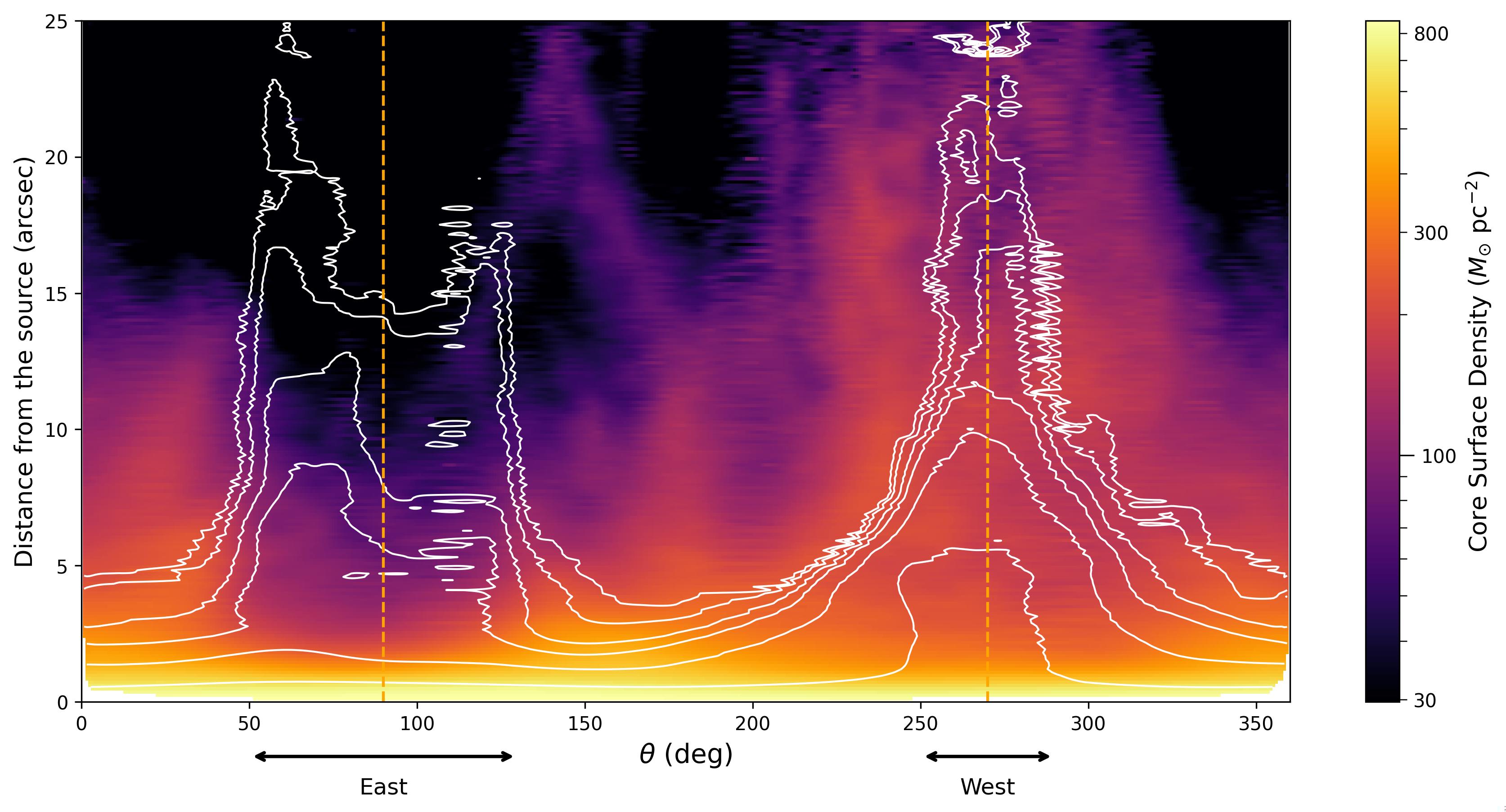}
    \caption{Surface density map of the core (image) and molecular outflow (white contours) at different distances away from the source and different relative angles. The color scale is in log scale and clipped below $30~ M_{\odot}~\mathrm{pc}^{-2}$. The angles are measured counterclockwise relative to the direction perpendicular to the outflow axis, which is indicated with $0^{\circ}$ in Figure \ref{fig:mass_core}. The contours for the molecular outflow surface density at levels of $5\%, 10\%, 20\%, 30\%, 50\%$, and $80\%$ of the maximum ($106~ M_{\odot}~\mathrm{pc}^{-2}$). The dashed orange lines indicate the directions along the outflow axis ($90^{\circ}$ and $270^{\circ}$), with the eastern and western lobe labeled below. 
    }
    \label{fig:mass_polar}
\end{figure*}

Evidence of interactions between the outflows and the core can be inferred from the comparison of their morphologies. The east side of the core exhibits a broad U-shaped morphology, as indicated by the contour level corresponding to $170~M_{\odot}~\mathrm{pc}^{-2}$ (see Figure~\ref{fig:mass_core}). The curvature is more apparent toward the north, with the northern half of the U-shaped feature closely following the outflow contours. This structure likely shows the northern cavity walls created by the outflow.  
Interaction between the outflow and the core is also visible in Figure \ref{fig:mass_polar},
which illustrates the surface density distributions of the core and the outflow in  polar coordinates. 
The eastern molecular outflow, centered on $\theta \sim 90^{\circ}$ in Figure \ref{fig:mass_polar}, coincides with a wide cavity in the core. The surface density in this cavity  at a distance of $\sim10^{\prime\prime}$  ($\sim $ 3860 au) from the protostar is as low as the surface density beyond the outer edge of the core ( $>15^{\prime\prime}$ from the protostar). The cavity in the western lobe is less obvious. There is a small drop in the surface density along the outflow axis ($\theta \sim 270^{\circ}$ ) starting at $r \sim 17^{\prime\prime}$ away from the protostar (see Figure  \ref{fig:mass_polar}). 
The core regions immediately next to the cavities (e.g., $\theta \approx 50^{\circ}$ and $\theta\approx 240 ^{\circ}$) have a higher surface density than the rest of the core. These are likely the cavity walls created by the outflows.

\subsubsection{Asymmetry in the opening angle and analytical model}
\label{sec:shape_out}

\begin{figure}
    \centering
    \includegraphics[width=0.98\linewidth]{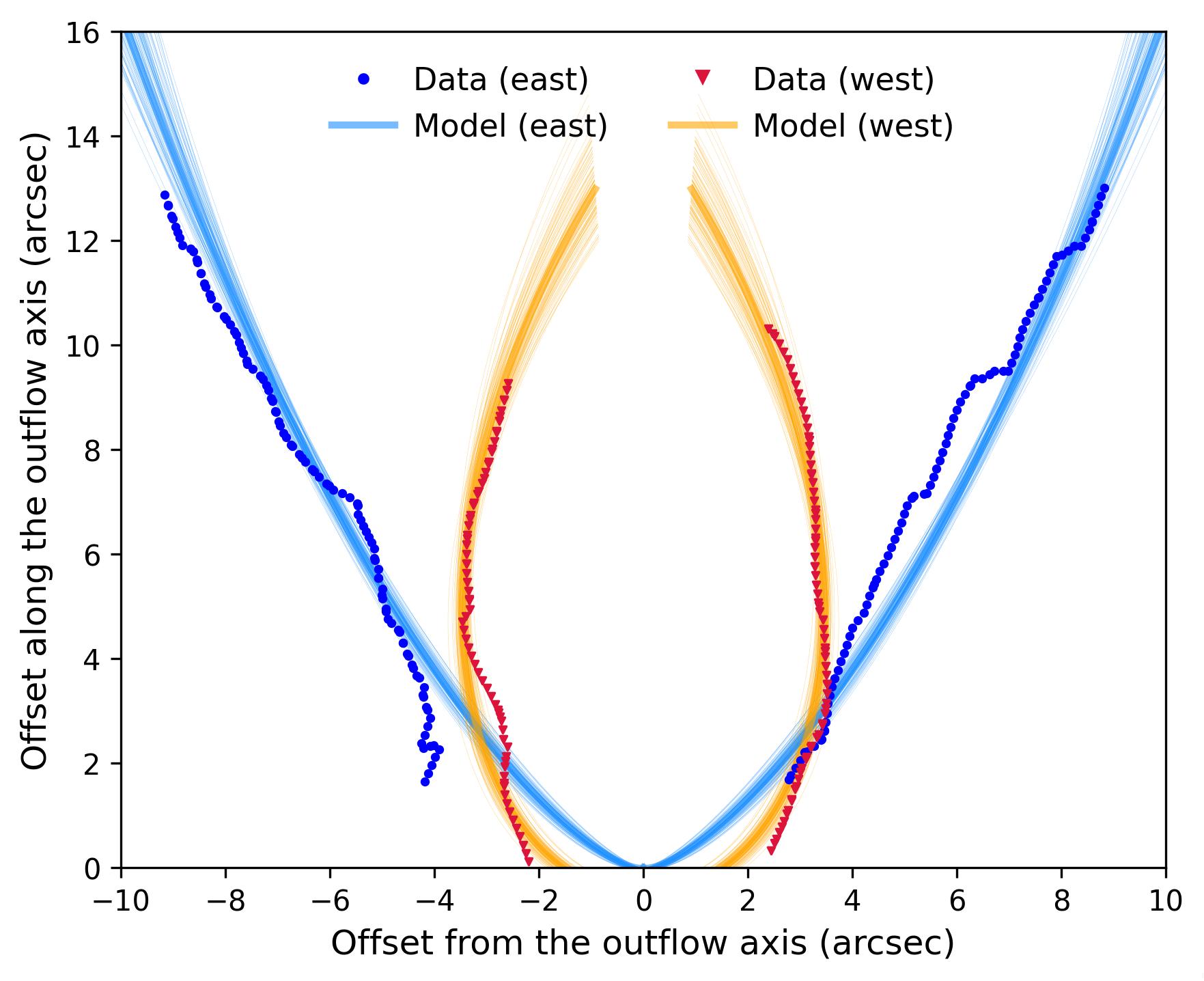}
    \caption{Shape of the outflow described by the analytical model discussed in Section \ref{sec:shape_out}.
    The data points trace the contours of the molecular outflow surface density map at $11~M_{\odot}~\mathrm{pc}^{-2}$ for the eastern lobe (blue) and $32~M_{\odot}~\mathrm{pc}^{-2}$ for the western lobe (crimson), corresponding to $10\%$ and $30\%$ of the maximum surface density (see the contours in Figures \ref{fig:mass_outflow} and \ref{fig:mass_polar}). The thick lines show the outflow shape given by Equation \ref{eq:shape_outflow2} using the median posterior parameters for  each lobe. The thin curves indicate outflow shapes using 100 random draws from the posterior distribution. 
    }
    \label{fig:shape_outflow}
\end{figure}

The difference in opening angles of the two lobes of the outflow is also striking in Figures \ref{fig:mass_outflow} and \ref{fig:mass_polar}. Figure \ref{fig:mass_polar} suggests that the eastern lobe of the outflow has the same opening angle of $80^{\circ}$ at different distances from the protostar. In contrast, the opening angle of the western lobe is much smaller and decreases with increasing distance from the protostar. The characteristic opening angle of the western lobe is estimated to be $30-40^{\circ}$ closer to the protostar and $\sim 20-30^{\circ}$ farther away.  

An asymmetry in opening angles like this was previously found in hydrodynamical simulations by \citet{Offner2011} and \citet{Offner2014}. Since these two simulations assumed symmetric winds launched from the protostar, the difference in opening angle was caused by the interactions between the wind and the core.  

It is thus natural to hypothesize that this is the case in HOPS 198 as well: the asymmetry in the opening angles of the eastern and western lobes of the outflow in HOPS 198 is caused by the asymmetry in the surface density of the core. As discussed above, the distribution of the core surface density with respect to the position of HOPS 198 suggests that this protostar is near the eastern edge of the core. 
As a result, the  symmetric wide-angle wind  launched  by the protostar encounters greater resistance in the western lobe where the gas is denser. This leads to a narrower opening angle in the western lobe than in the eastern lobe.  

We investigated the plausibility of this hypothesis further by developing an analytical model for the outflow morphology, building upon the work of \citet{Li2013}.  \citet{Li2013} hypothesized that the shape of the outflow cavities is determined by the balance between ram pressure of the protostellar wind and turbulent pressure of the core perpendicular to the outflow cavity wall. 

In a cylindrical coordinate system with the center at the position of the source and the $z$-axis pointing along the outflow axis, following \citet{Li2013}, the  shape of the cavity is given by 
\begin{equation}
\label{eq:shape_outflow1}
   \frac{\mathrm{d}z}{\mathrm{d}r_{\perp}} = \tan \left(\arctan \frac{z}{r_{\perp}} + \arcsin \frac{p_{\text{core}}}{p_{\text{wind}}} \right),
\end{equation}
where $z$ is  the distance along the outflow axis, $r_{\perp}$ is the distance from the outflow axis, and $p_{\text{core}}$ and $p_{\text{wind}}$ are the pressures of the core and the wind, respectively. 
This equation assumes that $p_{\text{core}} \leq p_{\text{wind}}$, which is reasonable given that the outflow cavity is stable. 
The pressures are given by
\begin{align}
    & p_{\text{wind}} = \rho_{\text{wind}} v_{\text{wind}}^2 \\
    & p_{\text{core}} = \rho_{\text{core}} \sigma_{\text{core}}^2,
\label{eq:pcore}
\end{align}
where $\rho_{\text{core}}$ and $\rho_{\text{wind}}$ are the volume densities of the gas in the core and the wind, respectively, $v_{\text{wind}}$ is the velocity of the  wind, and $\sigma_{\text{core}}$ is the gas velocity dispersion in the core. 

\citet{Li2013} derived Equation \ref{eq:shape_outflow1} assuming that the density distribution of the wind takes the form
\begin{equation}
\label{eq:rho_wind_Li}
    \rho_{\rm wind} \sim \frac{1}{r_{\rm sph}^2 \sin^2 \theta_{\rm sph}},
\end{equation}
where $r_{\rm sph}$ and $\theta_{\rm sph}$ are distance from the center and azimuthal angle in spherical coordinates. This form is a reasonably good approximation to radial hydromagnetic winds that have expanded to large distances, where the magnetic field becomes approximately force-free \citep{Ostriker1997, Matzner1999}. This approximation holds independent of the wind launching mechanisms and applies to X-winds \citep{Shu1994, Shu1995} and disk winds \citep{Pudritz2007}.

We argue that Equation \ref{eq:shape_outflow1} can be applied to more general winds when two conditions are satisfied: 1) the scale of the modeled outflow is significantly larger than the region from which the winds are launched, and 2) the opening angle of the winds at launch is larger than the opening angle of the cavity. Under these conditions, the winds can exert a pressure perpendicular to the cavity walls with the appropriate geometry for Equation \ref{eq:shape_outflow1} to remain valid.

\citet{Li2013} simulated outflows numerically using Equation \ref{eq:shape_outflow1} after adopting Equation \ref{eq:rho_wind_Li} and making assumptions about the mass-loss rate for the wind,
the density profile of the core, and the turbulent velocity of the core. 
To facilitate the comparison with observations and be less model-dependent, while we built upon Equation \ref{eq:shape_outflow1}, we did not make such assumptions. Instead, we assumed that the ratio of pressures $\xi = \frac{p_{\text{core}}}{p_{\text{wind}}}$ is roughly constant (see the paragraphs below for a discussion of the implications and limitations of this assumption). This assumption allowed us to rewrite Equation \ref{eq:shape_outflow1} as
\begin{equation}
    \frac{\mathrm{d}z}{\mathrm{d}r_{\perp}} = \frac{\frac{z}{r_{\perp}} + \alpha}{1- \alpha \frac{z}{r_{\perp}}},
\end{equation}
where $\alpha = \frac{\xi}{\sqrt{1-\xi^2}}$. The solution to this differential equation is the implicit equation
\begin{multline}
\label{eq:shape_outflow2}
    F(r_{\perp}, z)  = \arctan \left(\frac{z}{r_{\perp}}\right) - \alpha \ln \left( \frac{r_{\perp}}{r_0} \right) \\ - \frac{\alpha}{2} \ln \left[ 1 + \bigg( \frac{z}{r_{\perp}} \bigg)^2 \right] = 0,     
\end{multline}
where $r_0$ is the width of the outflow at the base ($z=0$). 

Equation \ref{eq:shape_outflow2} is governed by two parameters, $\alpha$ and $r_0$. $\alpha=\frac{\xi}{\sqrt{1-\xi^2}}$ is related to the ratio of the core turbulent pressure and the wind ram pressure, $\xi$. A larger $\alpha$, which corresponds to a larger $\xi$ and a stronger turbulent pressure in the core, leads to a narrower outflow cavity. On the other hand, $r_0$ controls the width of the outflow at the base. Under a fixed $\alpha$, a larger $r_0$ leads to a wider opening angle at a given distance along the outflow axis. 

\subsubsection{Implications of the analytical model for HOPS 198}
The shape of each lobe of the outflow, given by the contours in Figure \ref{fig:mass_outflow}, can be described and fit with Equation \ref{eq:shape_outflow2}. Since the inclination of the HOPS 198 system with respect to the plane of the sky is nearly edge-on \citep[$i\sim80^{\circ}$,][]{Hsieh2023}, 
any inclination correction would not alter the outflow morphology significantly, and we thus did not apply such corrections.

We adopted contour levels of $11~M_{\odot}~\mathrm{pc}^{-2}$ for the eastern lobe of the outflow and $32~M_{\odot}~\mathrm{pc}^{-2}$ for the western lobe. These values correspond to $10\%$ and $30\%$ of the maximum surface density, as shown in Figures \ref{fig:mass_outflow} and \ref{fig:mass_polar}. The contours capture the shape of the outflow reasonably well and encompass
$\sim 70\%$ of the mass in each lobe.
Parts of the contours that are too close to the protostar were removed because they are subject to greater uncertainties due to potential contamination from the inner envelope. 
The fit to the implicit equation was then carried out using a Markov chain Monte Carlo (MCMC) method using the \texttt{emcee} package \citep{ForemanMackey2013}, with the following likelihood function:
\begin{equation}
    \mathrm{log~likelihood} = -\frac{1}{2}\sum_{i} \left(\frac{F(r_{\perp}^{(i)}, z^{(i)})}{|\nabla F(r_{\perp}^{(i)}, z^{(i)})|} \right)^2,
\end{equation}
where $F(r_{\perp}, z)$ is the implicit function whose zeros define the curve in Equation \ref{eq:shape_outflow2}, $\nabla F(r_{\perp}, z)$ is the gradient of the function, and $(r_{\perp}^{(i)}, z^{(i)})$ are the coordinates of the points on the contours. This likelihood function was adopted from the cost function commonly used for fitting curves defined by the implicit equations from \citet{Taubin1991}. 

Figure \ref{fig:shape_outflow} shows the outflow shape given by Equation \ref{eq:shape_outflow2} using the medians of the posterior distributions of the MCMC fit of the parameters and 100 random draws from the posterior distributions. Figure \ref{fig:mass_outflow} also shows the curves given by the median posterior model along with the molecular outflow mass map.
The analytical equation is able to describe the shape of the two lobes reasonably well and is relatively robust, even though we made the highly simplistic assumption that $\xi$ is constant. 

The ratio of the core turbulent pressure to the outflow pressure inferred from the MCMC fit is $\xi=0.20^{+0.02}_{-0.02}$ and $\xi=0.58^{+0.03}_{-0.03}$ for the eastern and western lobe, respectively, and the $r_0$ we inferred is $0.12^{+0.08}_{-0.06}~^{\prime\prime}$ and $1.56^{+0.18}_{-0.18}~^{\prime\prime}$ for the eastern and western lobe respectively.   

The derived values of $r_0$ are not reliable estimates of the true width of of the outflow lobes at the base because of the oversimplifying assumption that $\xi$ is constant throughout the system. 
In reality, the density of the inner envelope ($\lesssim1000~\mathrm{au}$) surrounding the protostar is expected to be significantly higher than the rest of the core, which can lead to very different values of $\xi$. Our estimates of $\xi$ are constrained by the morphology of the extended outflows and are unlikely to be representative of the inner envelope. Considering the degeneracy between $r_0$ and $\xi$, as discussed in the previous paragraphs, the inferred values of $r_0$ should be treated with caution.

Moreover, the oversimplifying assumption is likely the reason for the large difference in $r_0$ between the two lobes. The volume density of the inner envelope should be approximately the same on two sides of the disk, as indicated by the almost identical surface density in the inner $4^{\prime\prime}$ of the core shown in Figure \ref{fig:density_asym}. The value of $\xi$ should thus be the same for the two sides in these regions. The significant underestimation of $\xi$ for the east inner envelope by our assumption that sets a global $\xi$ has to be compensated for by a smaller $r_0$, leading to a large difference in $r_0$ between the two lobes. 

However, this difference in $r_0$ between the two lobes does not explain the difference in opening angle, as a larger $r_0$ should lead to a wider outflow,  but the opposite is observed in HOPS 198. From our analysis, the difference in the opening angles of the fits to each lobe in this source is determined by the difference in $\xi$.

As indicated above, the ratio of the turbulent pressure to outflow pressure $\xi$ is given by
\begin{equation}
    \xi = \frac{\rho_{\text{core}} \sigma_{\text{core}}^2}{\rho_{\text{wind}} v_{\text{wind}}^2 }.
\end{equation}
If symmetric winds are launched from the protostar, the density and velocity of the wind should be the same for the two lobes. We find no compelling evidence of a significant difference in the turbulence level for the two sides of the core. The velocity dispersion differs by less than $30\%$ between the two sides of the core, based on the full width at half maximum of the C$^{18}$O spectra from circular apertures of radius $1.5^{\prime\prime}$ outside the outflow lobes. 
Therefore, the difference of factor of $\sim2.9$ in the pressure ratio $\xi$ is most likely caused by a difference in the gas density between the two sides of the core. That is,  the west side of the core would have to be denser by about  $2.9$ times than the east side. This interpretation is consistent with our independent result, where  we found that the surface density of the west core side is $1.5-2.8$ times higher than that of the east side (see Figure \ref{fig:density_asym}). 
The small discrepancy between these two ratios can be explained by the simplifying assumption of a global constant $\xi$ and by the need to deproject the surface density into volume density, a model-dependent process that requires assumptions about the underlying density profile.
Therefore, it is relatively safe to conclude that the difference in opening angle of the two lobes is caused by the difference in the density between the two sides of the core.

As discussed in the previous paragraphs, Equation \ref{eq:shape_outflow1}, and thus, the model we developed, assumes that the opening angle of the protostellar wind at launch is wider than the opening angle of the cavity.
When the opening angle of the cavity reaches the opening angle of the wind at launch, the wind can no longer exert any pressure perpendicular to the cavity wall, and the cavity does not grow further. The cavity opening angle on the east side is $\sim80^{\circ}$, suggesting that the wind is at least $80^{\circ}$ wide when launched. 

One limitation of our analytical model is the assumption that $\xi$ remains constant. This simplification is necessary because the available data do not allow us to robustly constrain $\xi$ as a function of distance from the protostar. For example, current observations do not have the resolution to estimate the column density of the core at disk scales, which is crucial for the deprojection of column density to the volume density of the core $\rho_{\rm core}$. Current observations also lack tracers of the launched protostellar winds, so $\rho_{\rm wind}$ cannot be constrained.

However, future observations with wind tracers at sufficiently high resolutions may allow us to estimate $\xi$ as a function of radius solely from observations. This can allow us to compute the shape of the outflow from Equation \ref{eq:shape_outflow1}, which can be compared with direct observations of the outflow. A direct comparison like this would help us to quantitatively bridge the gap between theory and observation for individual protostellar systems, which has been difficult to achieve for simulation studies. 

In a recent study, \citet{Rabenanahary2022} modeled and simulated molecular outflows driven purely by pulsed jets. Their model predicted outflows similar to those driven by wide-angle winds and observed outflows in systems such as HH46-47. 
The opening angle of the jet-driven outflows is limited by the deceleration of jet-driven shells and does not grow significantly after $\sim8000$ yr. This contrasts with our model of outflows entrained by a wide-angle wind, in which the outflow opening angle is ultimately limited by the opening angle of the wind at launch. \citet{Rabenanahary2022} also found that the outflow opening angle for a jet-driven outflow is generally more narrow than a wind-driven outflow over time (see their Figure 6). 
A large sample of protostars with a morphology observed at high resolution might help us to distinguish between the two models. Their simulations, however, showed that the opening angle of outflows driven by pulsed jets is also sensitive to the density of the protostellar core, like the models presented in this paper. Thus, our interpretation of the asymmetries in HOPS 198 is still reasonable even when the outflows are jet-driven.

\subsection{Implications for the evolution of outflow opening angles}
\label{sec:oa_evol}

As discussed in Section \ref{sec:intro}, while the increase in the protostellar outflow opening angle over time is well established, the mechanism for this increase is not clearly understood. A reasonable explanation was proposed by \citet{Arce2006}, who suggested that a collimated jet and a wide-angle wind coexist in a protostellar system. As the core and envelope lose mass, the wide-angle wind can break through the surrounding material, allowing the outflow cavity to widen. 

Our analysis of the asymmetric protostellar system HOPS 198 provides direct observational evidence that the outflow opening angle is strongly affected by the density of the surrounding core. 
Our results offer a quantitative test of the interpretation proposed by \citet{Arce2006} that the widening of outflow cavities over time is linked to the outflow-core interactions. Our analytical model and quantitative results advance this picture and yield a more refined dynamic view of outflow evolution, in which changes in the core density regulate the extent of the  interaction between an intrinsically wide-angle protostellar wind and  the ambient medium. 

This dynamic picture can be described as follows. The initial high density of the envelope creates a strong turbulent pressure that successfully opposes the ram pressure from the wide-angle wind and limits the width of the outflow cavity. As core material is removed by the outflow and accreted onto the protostar, the density of the envelope and  core decrease and the strength of the ambient turbulent pressure declines. If the ram pressure from the wind remains strong, the outflow cavities widen over time. This can explain the widening of outflow cavities  during the evolution of  protostars from Class 0 to Class I. 

Eventually, when the cavity opening angle reaches the angle of the protostellar wind at launch, the wind no longer exerts pressure perpendicular to the cavity wall. Therefore, the opening angle no longer grows. This might explain the observed decrease in the rate at which outflows widen for more evolved Class I and flat-spectrum protostars. 
Depending on the initial cloud environment and the properties of the launched wide-angle wind, the evolutionary timescales of the cavity opening angle can be different for different protostars, which explains the scatter in the opening angle for protostars with the same bolometric temperature as observed by \citet{Dunham2024}.
In certain systems, such as HOPS 198, the initial density gradient in the cloud allows the two lobes to evolve under different timescales, which leads to the observed asymmetric outflow lobes. 

Our interpretation is partly supported by studies of the evolution of envelope densities. \citet{Furlan2016} presented fits to the spectral energy distributions of 330 sources in the Herschel Orion Protostar Survey and found that the median envelope density decreased from Class 0 to Class I to flat-spectrum  II sources. This result was confirmed by a follow-up study by \citet{Pokhrel2023}. However, both studies assumed that the core density profile follows a power law, which might oversimplify the situation and result in biases. 
At the same time, our results do not rule out the possibility that other factors affecting wind-core interaction also contribute to the evolution of the opening angle. For example, \citet{Li2013} proposed that the evolution in the strength of protostellar winds and turbulence of the core, instead of density of the core, might cause the changes in the outflow opening angle.

\section{Conclusions}

We examined the asymmetric core and molecular (CO) outflow of the protostellar system HOPS 198 using ALMA Band 6 observations.

We found that the surface density in the east side of the core is roughly twice lower than the west side. On the other hand, the eastern lobe of the outflow (with an opening angle of $\sim80^{\circ}$) is much wider than the western lobe (with an opening angle of $\sim30^{\circ}$). Building upon \citet{Li2013}, we developed an analytical model that describes the shape of the outflow cavity using $\xi$, the ratio of turbulent pressure from the core and the ram pressure from the protostellar wind. We found that the difference in opening angle of the two molecular outflow lobes can be explained by the difference in density, which leads to a difference in turbulent pressure in the two sides of the core.  

This result provides direct evidence that the density of the core determines the molecular outflow opening angle. Based on this result, we propose that the increase in the outflow opening angle as protostars evolve might be a dynamic process, as discussed by \citet{Arce2006}. The opening angles of molecular outflow  lobes widen over time as the density in the core decreases through accretion and gas dispersal. When the outflow opening angle reaches the opening angle of the underlying protostellar wind at launch, the outflow cavity opening angle stops growing and remains constant. This dynamic picture is consistent with the observed evolution of outflow opening angle of protostars at different evolutionary stages  (e.g., \citealt{Hsieh2023, Dunham2024}). 

While our interpretation provides a refined explanation for the evolution in the protostellar outflow and is indirectly supported by the decrease in the ambient envelope density over time \citep{Furlan2016, Pokhrel2023}, future observations of the environment around a large sample of protostars are necessary to confirm it. Moreover, the implications of our finding on the core gas dispersal mechanism by outflows and its effect on the mass-assembling process should be further investigated.

\begin{acknowledgements}
We thank the anonymous referee for their constructive suggestions.
D.W. acknowledges support from Yale College First-Year Summer Research Fellowship in the Sciences $\&$ Engineering. 
This work was also supported by NSF grant AST-2407116 awarded to H.G.A.
C.H.H. is supported by the NASA Hubble Fellowship Program under award HST-HF2-51556.

This paper makes use of the following ALMA data: ADS/JAO.ALMA$\#$2018.1.00744.S. ALMA is a partnership of ESO (representing its member states), NSF (USA) and NINS (Japan), together with NRC (Canada), NSTC and ASIAA (Taiwan), and KASI (Republic of Korea), in cooperation with the Republic of Chile. The Joint ALMA Observatory is operated by ESO, AUI/NRAO and NAOJ.
\end{acknowledgements}

\bibliographystyle{aa}
\bibliography{HOPS198}

\FloatBarrier

\begin{appendix}
\nolinenumbers

\section{Column density maps from Herschel and C$^{18}$O emission}
\label{sec:Herschel}

\begin{figure*}[!t]
    \sidecaption
    \includegraphics[width=12cm]{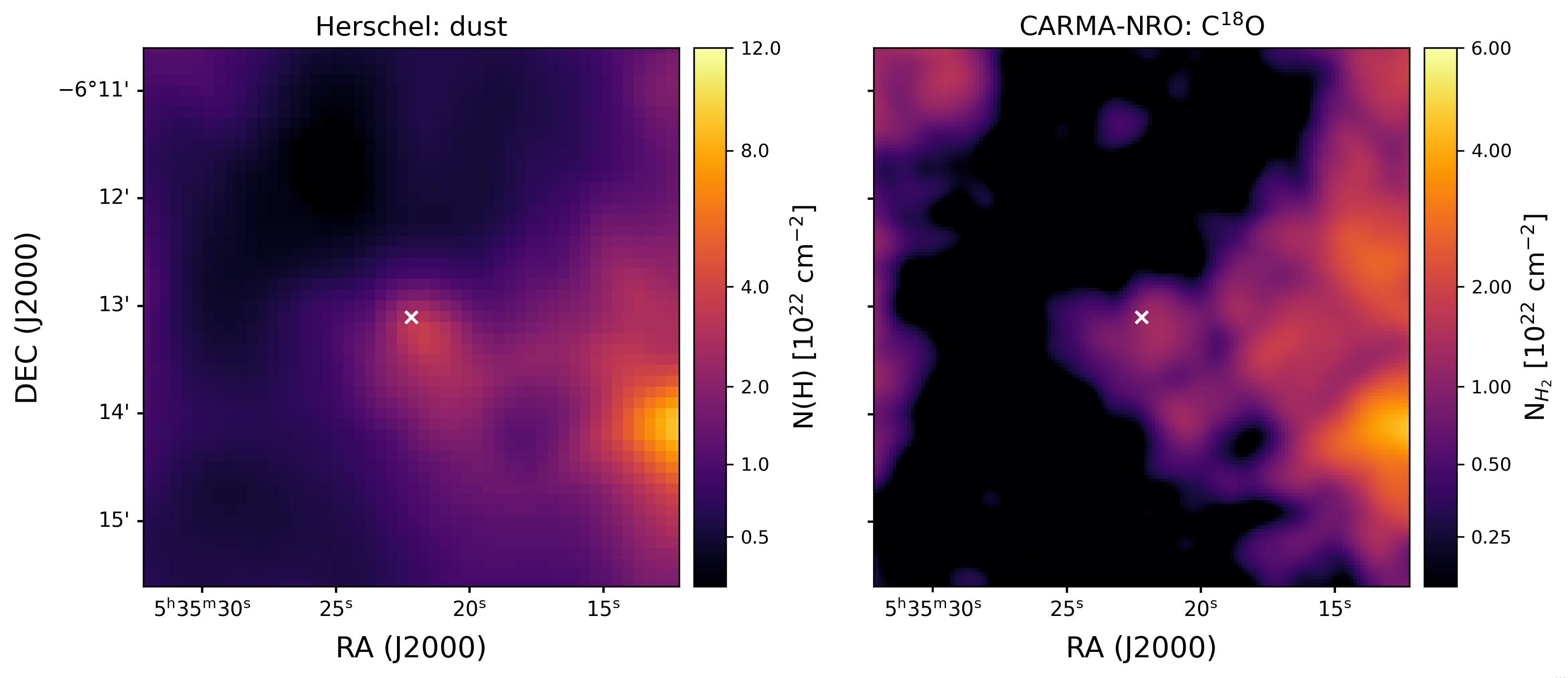}
    \caption{Left: Total hydrogen column density ($N(H)$) around HOPS 198 derived by \citet{Stutz2015} using dust emission observed by Herschel. Right: The column density of molecular hydrogen H$_2$ around HOPS 198 derived using C$^{18}$O (1-0) emission observed by the CARMA-NRO Orion survey  (as described in Appendix \ref{sec:mass}), convolved with a $18^{\prime\prime}$ beam. The cross indicates the position of HOPS 198.
     }
    \label{fig:herschel}
\end{figure*}

Low-$J$ transitions of C$^{18}$O have been used as tracers for protostellar cores for decades \citep{Myers1983, Zhou1994, Onishi1996, Arce2006}. Their intensities offer relatively robust estimates of column densities, as they remain mostly optically thin on core scales and correlate well with column densities estimated from dust emission, except for cold cores with very high visual extinction \citep{Alves1999, Pineda2008}.

To validate our analysis of the core surrounding HOPS 198 using C$^{18}$O, we compare column densities derived from C$^{18}$O and dust continuum emissions.
Dust emission in Orion A has been observed using \textit{Herschel} at 160, 250, 350, and 500 $\mathrm{\mu m}$ as part of the Herschel Gould Belt program \citep{Polychroni2013}. By modeling the dust emission using a modified blackbody spectrum, \citet{Stutz2015} have derived the total hydrogen column density in Orion A, $N(H)$. The left panel of Figure \ref{fig:herschel} shows the total hydrogen column density around HOPS 198. The right panel of Figure \ref{fig:herschel} shows the H$_2$ column density around HOPS 198, estimated using CARMA-NRO observations (see Appendix \ref{sec:mass}). To compare the two maps, we convolve the CARMA-NRO map with a Gaussian beam with an FWHM of $18^{\prime\prime}$, matching the beam size of the 250 $\mu\mathrm{m}$ intensity map used to estimate $N(H)$. The CARMA-NRO map still appears to have better resolution, because the temperature map used in deriving $N(H)$ has a larger beam (FWHM = $36^{\prime\prime}$, \citealt{Stutz2015}).

Comparison between the two panels show that the morphology of the core deduced from dust emission is consistent with that deduced from C$^{18}$O emission. The column densities estimated are also consistent between the two tracers: $N(H)$ estimated from dust emission is roughly $2$ times higher than H$_2$ column density from C$^{18}$O emission. This is expected as $N(H) = N_{HI} + 2N_{H_2}$, where $N_{HI}$ is column density of atomic hydrogen and is negligible in molecular clouds like Orion.
The consistency in the morphology and values of column densities with dust emission observed by Herschel suggests that C$^{18}$O emission does trace protostellar cores. This validates our methodology.

\section{Mass of the core and the outflow}
\label{sec:mass}
Following \citet{Dunham2014}, the column density of CO isotopologs in a given pixel can be computed from optically thin molecular line emission using:
\begin{equation}
\label{eq:N_CO}
    N_{\text{CO}} = \frac{3k}{8\pi^3 \nu \mu^2} \frac{(2J+1)}{(J+1)} \frac{Q(T_{\text{ex}})}{g_J} e^{\frac{E_{J+1}}{kT_{\text{ex}}}} \int T_{\text{mb}}(v_{\text{los}}) \, \mathrm{d}v_{\text{los}},
\end{equation}
where $k$ is the Boltzmann constant, $J$ is the rotational quantum number of the lower state in the transition from state $J+1$ to $J$, $g_J$ is the statistical weight or degeneracy of state ($J$), and $T_{\text{mb}}(v_{\text{los}})$ is the brightness temperature of the emission of the molecular transition in a channel with line-of-sight velocity $v_{\text{los}}$.  The frequency ($\nu$), magnetic dipole moment ($\mu$), and upper-level energy ($E_{J+1}$)  are specific to the molecular transitions of the CO isotopologs. $T_{\text{ex}}$ is the excitation temperature of the transitions. We assume that the CO isotopologs are in local thermal equilibrium at the same excitation temperature of $T_{\text{ex}} = 20$ K. 

The column density of H$_2$ gas for a given pixel can be derived from the column density of CO isotopologs assuming an abundance ratio:
\begin{equation}
    N_{\text{H}_2} = \frac{N_{\text{CO}}}{X_{\text{CO, H}_2}},
\end{equation}
where $X_{\text{CO, H}_2} = N_{\text{CO}}/N_{\text{H}_2}$ is the abundance ratio of the CO isotopolog relative to H$_2$. We assume an abundance ratio of $X_{^{12}\text{CO, H}_2} = 10^{-4}$ for $^{12}$CO. With $N_{^{12}\text{CO}}/N_{^{13}\text{CO}} = 62$ \citep{Langer1993} and $N_{^{12}\text{CO}}/N_{\text{C$^{18}$O}} = 540$ \citep{Wilson1992}, the abundance ratios for $^{13}$CO and C$^{18}$O are $X_{^{13}\text{CO, H}_2} = 1.61\times10^{-6}$ and $X_{\text{C$^{18}$O, H}_2} = 1.85\times10^{-7}$. 

The mass per pixel is then easily derived from the H$_2$ column density \citep{Hsieh2023}:
\begin{equation}
\label{eq:mass}
    M = \mu_{\text{H}_2} m_\text{H} A_{\text{pixel}} N_{\text{H}_2},
\end{equation}
where $\mu_{\text{H}_2} \approx 2.8$ is the mean molecular weight of H$_2$ \citep{Kauffmann2008}, $m_\text{H}$ is the hydrogen atom mass, and $A_{\text{pixel}}$ is the area of each pixel. The surface density is given by $\Sigma= M / A_{\text{pixel}}$.

\subsection{Mass of the core}
We estimated the mass of the core primarily using the molecular line C$^{18}$O (2-1) observed by ALMA, as it traces relatively dense gas associated with the core of HOPS 198, as discussed in Appendix \ref{sec:Herschel}. We assume that the emission is optically thin. Under this assumption, Equations \ref{eq:N_CO} to \ref{eq:mass} can be applied directly to the C$^{18}$O emission to compute the the mass of the core. 

We applied a $3\sigma$ mask to the C$^{18}$O emission to avoid contributions from noise when we compute the integrated intensity in Equation \ref{eq:N_CO}. Since there is negligible  contamination from outflows or other cloud components in the C$^{18}$O(2-1), all velocity channels are included in  core mass estimate. 

We also computed the mass and surface density of the core on a larger scale using the C$^{18}$O (1-0) emission, observed by the CARMA-NRO Orion survey \citep{Kong2018}, which we assume to be optically thin.  Since cloud emission not associated with the HOPS 198 core is visible in the CARMA-NRO map on a scale larger than our ALMA map field of view, only channels $4.11 \leq v_{\text{los}} \leq 7.19$ km/s are included to reduce possible contamination from other cloud components. Like for the ALMA observations, a $3\sigma$ mask is applied.

\subsection{Mass of the outflow}

\begin{figure*}
    \sidecaption
    \includegraphics[width=12cm]{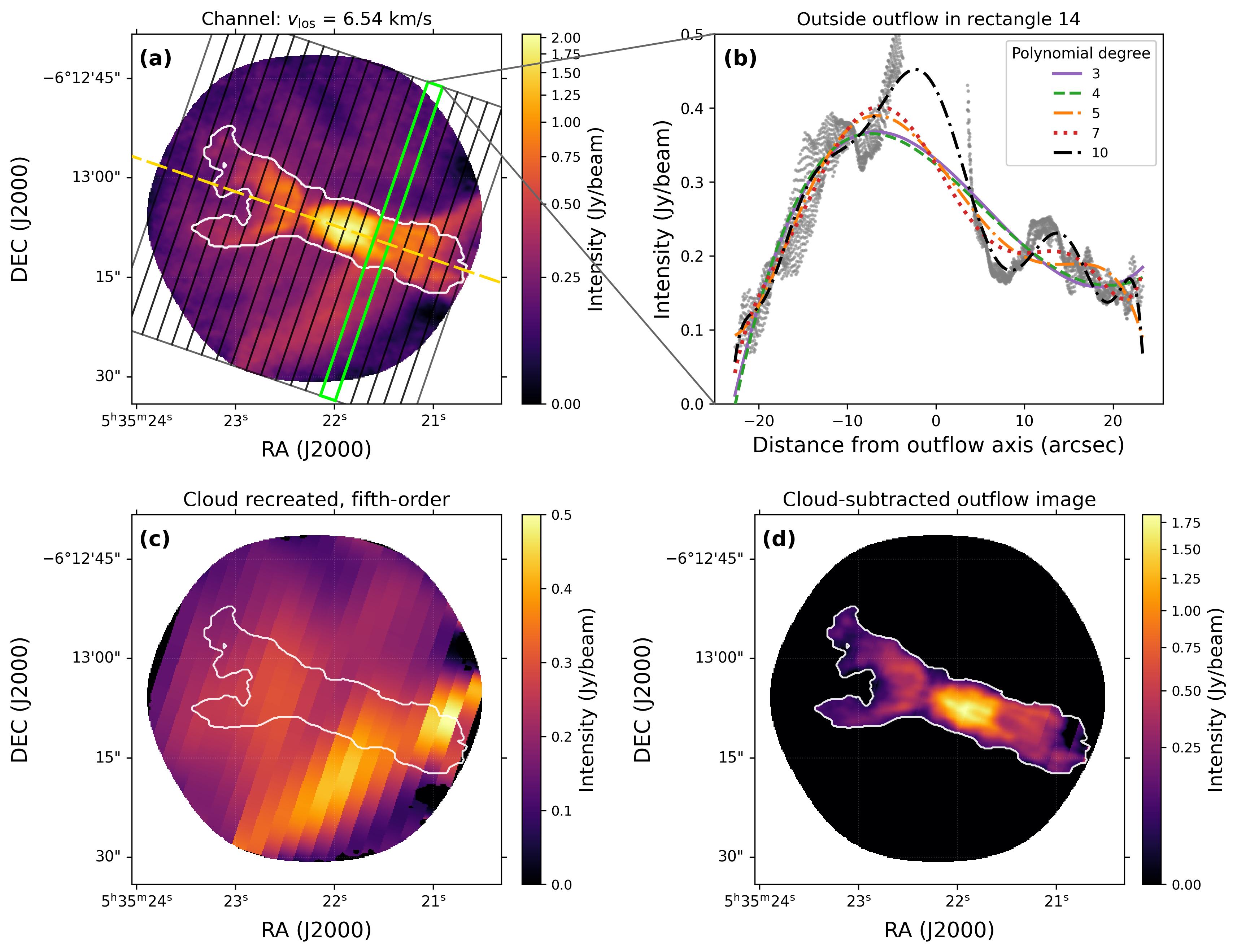}
    \caption{a) The $^{12}$CO emission in the channel with $v_{\rm los}=6.54$ km/s. The color scale uses a square-root stretch. The white contour indicates the outflow mask used for this channel, and the gold dashed line indicates the outflow axis. The black rectangles represent the rectangles created for the cloud subtraction method, while the lime rectangle highlights the specific rectangle for which the intensity profile is shown in panel (b). b) The intensity as a function of distance from the outflow axis for every pixel outside the outflow mask for the lime rectangle highlighted in panel (a). The lines show the best-fit polynomials with different degrees (3, 4, 5, 7 and 10) to the intensity-distance relation. c) The $^{12}$CO emission from the cloud recreated from the best-fit fifth-order polynomial. d) The image of the outflow inside the outflow mask after cloud subtraction. The color scale has a square-root stretch.
    }
    \label{fig:cloud_sub}
\end{figure*}

$^{12}$CO (2-1) line traces  the molecular outflow;  the cloud material entrained by the protostellar wind. Computing the mass of the outflow from $^{12}$CO emission presents two challenges: 1) there is significant $^{12}$CO emission from the core and cloud, especially around the systemic velocity; 2) the $^{12}$CO emission is considerably optically thick near the systemic velocity. 

\begin{figure*}[!b]
    \centering
    \includegraphics[width=0.9\textwidth]{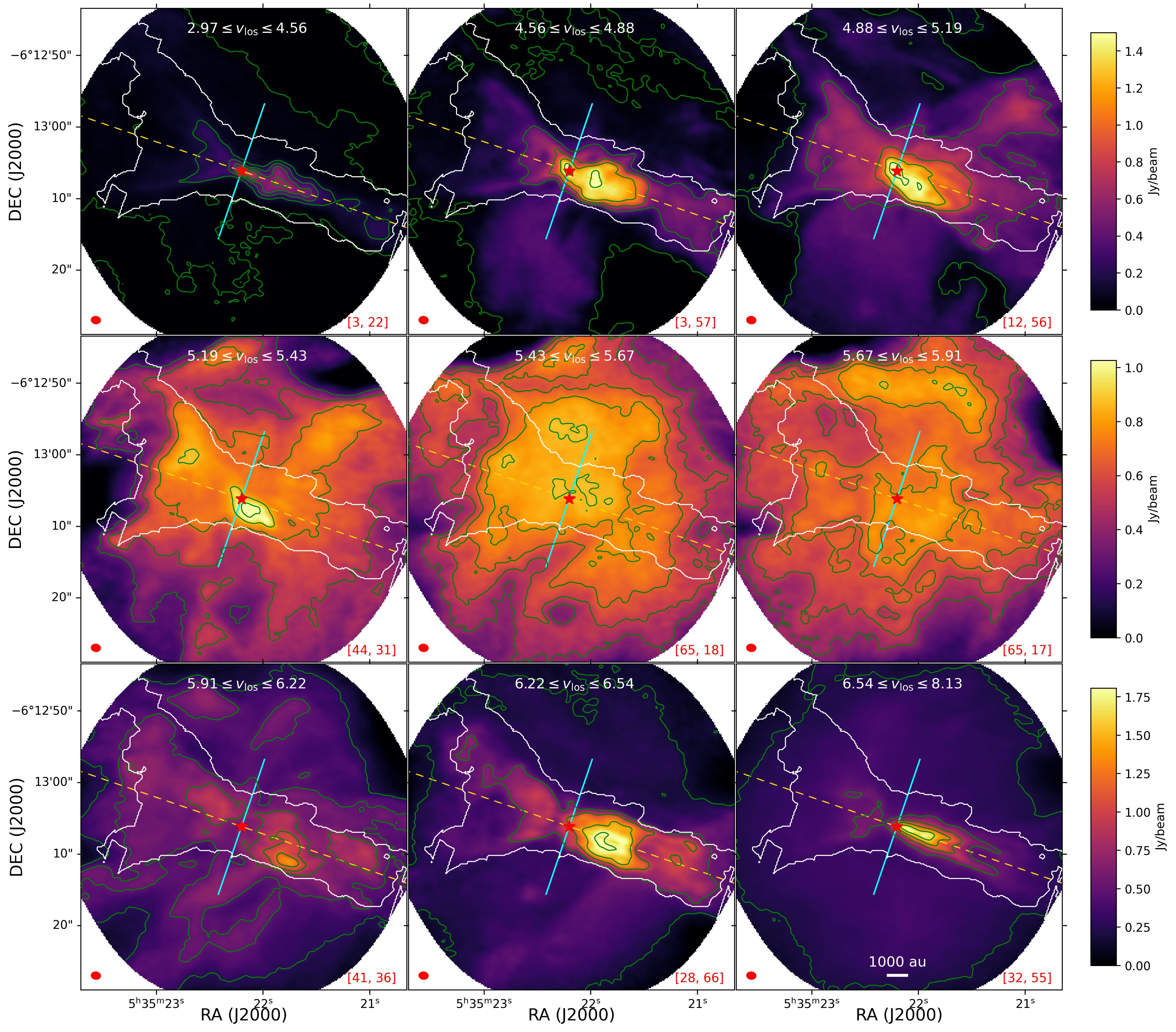}
    \caption{$^{12}$CO (2-1) emission, averaged over velocity channels with central velocities in km/s labeled at the top of each panel. All panels within a given row share a common color scale, capped at 99.9th percentile of pixel intensities across that row. The top and bottom rows are shown with an arcsinh stretch, and the middle row is shown with a linear scale. The green contours represent the  $^{12}$CO integrated intensity in multiples  of $\sigma$s (rms noise in the center of the image, 6.3 $\mathrm{mJy\,beam^{-1}}$). The starting contour and contour steps, in multiples 
     of $\sigma$s are indicated in brackets in the bottom right corner of each panel. The white contour shows the $3\%$ of the maximum value of the molecular outflow surface density ($106 ~M_{\odot}~\mathrm{pc}^{-2}$). The dashed gold line indicates the outflow axis, and the solid cyan line indicates the direction perpendicular to the outflow axis. The red star shows the location of the protostar. The red ellipse in the left bottom corner of each panel indicates the size of the beam. The physical scale is indicated in the final panel. 
     }
    \label{fig:12CO_channel_map}
\end{figure*}

To isolate the outflow emission as much as possible, without incorporating contaminating emission from the core and cloud, we implemented three techniques: exclusion of velocity channels, limiting the outflow mass estimate to regions enclosed within determined ``outflow masks'', and cloud emission subtraction for emission inside these masks. 
We exclude channels $5.19 \leq v_{\text{los}} \leq 6.07$ km/s and channels $8.45 \leq v_{\text{los}} \leq 10.19$ km/s, from the outflow mass estimate as in these velocities emission from various cloud components dominate over the outflow emission (see Figure \ref{fig:12CO_channel_map}). 
 
We produced several different masks, depending on the velocity channel, as the outflow morphology changes with velocity. Four outflow masks are created using contours in the integrated intensity maps, integrating over the following velocity ranges: $4.16 \leq v_{\text{los}} \leq 5.11$ km/s; $6.15 \leq v_{\text{los}} \leq 7.02$ km/s; $7.10 \leq v_{\text{los}} \leq 8.37$ km/s; and $10.27 \leq v_{\text{los}} \leq 10.75$ km/s. 
For channels $v_{\text{los}} < 4.16$ km/s and $v_{\text{los}} > 10.75$ km/s, the emission is almost entirely from the outflow. We applied a simple mask consisting of two triangles to each outflow lobe for $v_{\text{los}} < 4.16$ km/s, and no mask was applied for $v_{\text{los}} > 10.75$ km/s.

At velocities close to systemic velocity the cloud emission can be comparable to or exceed the emission from the molecular outflow lobes (even within the outflow masks), and thus cloud subtraction in the outflow masks is necessary. \citet{Hsieh2023} estimated the background emission in circular annuli centered at the protostar. However, this simple method 
does not take into account 
the highly asymmetric core and the localized emission structures present in HOPS 198.

Here we use an alternative method of cloud subtraction. In this method, the image in every channel is divided into 20 rectangles with the long side perpendicular to the outflow axis, each with length of 50$^{\prime\prime}$ and width of roughly 2.5$^{\prime\prime}$ (see panel (a) of Figure \ref{fig:cloud_sub}). For each rectangle, we find the intensity of the emission as a function of distance to the outflow axis for every pixel outside the outflow mask for that channel. We then fit a polynomial to the intensity-distance relation. Panel (b) of Figure \ref{fig:cloud_sub} shows an example of such fit. We assume that the estimated cloud emission in pixels inside the outflow mask is given by the value of the fitted polynomial. This is then subtracted to obtain outflow maps approximately free from cloud emission. 
We chose a fifth-order polynomial to balance between: 1) overfitting pixels that are right outside the outflow mask and still contain emissions from the outflow (see, for example, the sharp increase in intensity close to outflow axis in panel (b) of Figure \ref{fig:cloud_sub}); 2) underfitting of small localized emission structures. 
We verified the capability of this method to estimate the asymmetric or non-homogeneous emissions from clouds, by comparing the spectral cube produced using this cloud-subtraction technique with the original cube. Localized and extended emission are reliably reproduced by this method. For example, as shown in panel (c) of Figure \ref{fig:cloud_sub}, the localized emission from another cloud in the southwest direction of the source is successfully recreated by our method.

To account for the optical depth of $^{12}$CO emissions, we follow the procedures of opacity correction of \citet{Dunham2014}, \citet{Zhang2016} and \citet{Hsieh2023}. This method assumes that $^{12}$CO, $^{13}$CO and C$^{18}$O are in local thermal equilibrium at the same excitation temperature  and C$^{18}$O is optically thin. $^{13}$CO is corrected for opacity using the ratio of brightness temperature of the $^{13}$CO and C$^{18}$O emission, followed by opacity correction of $^{12}$CO using the  ratio of the brightness temperature of $^{12}$CO and $^{13}$CO emission. To reduce noise, the ratio of brightness temperature is averaged over outflow masks for each velocity channel and then fit with a second-order polynomial centered at the systemic velocity $v_{\rm sys}$. 

To account for the asymmetry of the system, we used two outflow masks for opacity correction, one for each lobe of the outflow. To reduce discontinuities at the interface of the two masks, an overlapping region centered on the protostar and extending 1$^{\prime\prime}$ parallel to the outflow axis was introduced. Linear interpolation of the correction factors was applied to the overlapping region. 

After cloud subtraction and opacity correction, the mass of the outflow can was estimated using Equation \ref{eq:mass}.

\section{$^{12}\mathrm{CO}(2\!-\!1)$ Emissions}
\label{sec:channel_map}

Figure \ref{fig:12CO_channel_map} shows the velocity-averaged channel maps of $^{12}$CO emission in HOPS 198 observed by ALMA.

\end{appendix}

\end{document}